\documentclass[3p]{elsarticle} 

\usepackage{lineno,hyperref}
\usepackage{lipsum}
\usepackage[export]{adjustbox} 

\usepackage{amsmath}
\usepackage{algorithmic}
\usepackage{array}
\usepackage{subfig}

\usepackage{url}
\usepackage{upgreek}
\usepackage{acronym}{}
\usepackage[normalem]{ulem}  

\usepackage{graphicx}
\usepackage{multirow}
\usepackage{booktabs}

\usepackage{caption}

\begin{document}

\begin{frontmatter}

\title{3D medical image segmentation with labeled and unlabeled data using autoencoders at the example of liver segmentation in CT images}
\author[1,2]{Cheryl Sital}
\author[2]{Tom Brosch}
\author[1,2]{Dominique Tio}
\author[1]{Alexander Raaijmakers}
\author[2]{J{\"u}rgen Weese\corref{cor1}}
\ead{juergen.weese@philips.com}

\cortext[cor1]{Corresponding author.}
\address[1]{Medical Image Analysis, Department Biomedical Engineering, Eindhoven University of Technology, Eindhoven, The Netherlands}
\address[2]{Philips GmbH Innovative Technologies, Hamburg, Germany}

\begin{abstract}
Automatic segmentation of anatomical structures with convolutional neural networks (CNNs) constitutes a large portion of research in medical image analysis. The majority of CNN-based methods rely on an abundance of labeled data for proper training. Labeled medical data is often scarce, but unlabeled data is more widely available. This necessitates approaches that go beyond traditional supervised learning and leverage unlabeled data for segmentation tasks. 
This work investigates the potential of autoencoder-extracted features to improve segmentation with a CNN.
Two strategies were considered. First, transfer learning where pretrained autoencoder features were used as initialization for the convolutional layers in the segmentation network. Second, multi-task learning where the tasks of segmentation and feature extraction, by means of input reconstruction, were learned and optimized simultaneously. 
A convolutional autoencoder was used to extract features from unlabeled data and a multi-scale, fully convolutional CNN was used to perform the target task of 3D liver segmentation in CT images. For both strategies, experiments were conducted with varying amounts of labeled and unlabeled training data.
The proposed learning strategies improved results in $75\%$ of the experiments compared to training from scratch and increased the dice score by up to $0.040$ and $0.024$ for a ratio of unlabeled to labeled training data of about $32 : 1$ and $12.5 : 1$, respectively. The results indicate that both training strategies are more effective with a large ratio of unlabeled to labeled training data.
\end{abstract}

\begin{keyword}
  CT liver segmentation \sep convolutional autoencoder \sep self-supervised learning \sep multi-task learning \sep transfer learning
\end{keyword}

\end{frontmatter}

\section{Introduction}

Deep learning applications addressing segmentation account for a vast amount of papers published in the field of medical image analysis \citep{litjens2017survey}. Segmentation of anatomical structures is an important step in radiological diagnostics and image-guided intervention, but expert manual segmentation of medical images, especially in 3D, is tedious and time-consuming. Automatic segmentation with convolutional neural networks (CNNs) presents an appealing alternative, as CNNs have achieved expert-level performance in certain applications. The U-Net architecture \citep{ronneberger2015u} is often considered as a benchmark in medical image segmentation, but the 3D variant can be challenging to train due to its high memory requirements. F-Net \citep{brosch2018foveal} is a 3D CNN which circumvents this memory issue by greatly decreasing the number of voxels that must be processed simultaneously. Furthermore, comparison experiments by \cite{brosch2018foveal} showed that the training for F-Net can be $4-10$ times faster than for an equivalent 3D U-Net architecture.

All of these CNN-based approaches require a large amount of labeled data to train effectively; they are supervised learning methods. However, labeled medical data is difficult to obtain since annotations are not always necessary in clinical practice and the collected annotations are often inaccessible to other researchers. This scarcity of labeled data is considered to be one of the biggest impediments to the success of CNNs \citep{litjens2017survey, shen2017deep}.
Larger quantities of unlabeled data are, in general, more readily available.

Unsupervised learning methods can learn to recognize objects and patterns without specific labels.
Autoencoders are one variant of unsupervised learning, which attempt to learn a reduced feature representation of the input data from which the original can be reconstructed. The subclass of regularized convolutional autoencoders (CAEs) has been shown to effectively extract features from unlabeled images for subsequent classification tasks \citep{makhzani2015winner, masci2011stacked}. In particular, the Winner-Take-All (WTA) CAE \citep{makhzani2015winner}, which enforces extreme sparsity during training, was proven to successfully learn shift-invariant image features from 2D non-medical images. 

In this study, we applied the WTA-CAE training method to unannotated medical volumes and used the extracted features to aid the target task of 3D segmentation. F-Net was selected as the network architecture for segmentation, due to the faster training and lower memory consumption. The question then arises of how to combine these two networks.

In their recent survey on not-fully-supervised learning techniques in medical imaging, \cite{cheplygina2019not} distinguish three types of learning scenarios.
Two are particularly relevant for segmentation: semi-supervised learning (SSL) and transfer learning (TL). 

SSL methods consider a set of unlabeled data in addition to the labeled training dataset to improve the classifier's predictions. Applications for segmentation include label propagation via self-training \citep{bai2017semi,meier2014patient} and co-training with additional classifiers \citep{iglesias2010agreement}, which attempt to classify unlabeled samples and add them to the training set. The performance strongly depends on the classification error for the unlabeled samples, since including wrongly classified training data can worsen results. Another possibility is graph-based methods\citep{mahapatra2016combining,song2009semi}, which try to link the distributions of the labeled and unlabeled data in order to regularize the classifier. 

TL aims to transfer knowledge from a related learning problem to the task at hand. The data for the related learning problem can come from a possibly different medical \citep{kushibar2019supervised, ghafoorian2017transfer} or a non-medical \citep{tajbakhsh2016convolutional, shin2016deep, menegola2017knowledge} domain. Despite some promising results with models pretrained on non-medical data, e.g. ImageNet \citep{russakovsky2015imagenet}, this approach restricts the choice of network architecture and is barely applicable to 3D tasks. An in-depth comparison of TL with medical and non-medical data has been presented by \cite{cheplygina2019cats}. 

Knowledge transfer from unlabeled medical data has been successfully performed for segmentation via pretraining of segmentation networks through self-supervised learning and multi-task learning. In self-supervised learning, labels are fabricated from unlabeled data without explicit human annotation to perform supervised tasks. These labels can be inherent to the data, for example using image metadata to predict anatomical positions \citep{bai2019self} and image re-colorization \citep{ross2018exploiting}. Another possibility is to modify the data appearance and try to predict the transformation, e.g. image rotation \citep{tajbakhsh2019surrogate}. In multi-task learning (MTL), additional unsupervised task(s) are performed, either sequentially \citep{dhungel2017deep, bai2019self} or simultaneously, in order to improve the supervised target task of segmentation. 

While many studies were encountered utilizing TL with unlabeled medical data for classification and diagnosis tasks, only a few cases were found adopting this approach for segmentation \citep{bai2019self,ross2018exploiting,tajbakhsh2019surrogate,blendowski2019how,chen2019self}.
This work investigates two strategies for combining knowledge from labeled and unlabeled data for segmentation. 
The first approach uses a WTA-CAE to learn features from unlabeled images and uses those features to initialize the weights of an F-Net that is then trained to perform the segmentation task.
This strategy can be likened to pretraining of the segmentation network with self-supervision and is hereafter denoted by S-TL.
The second approach simultaneously trains an F-Net for segmentation and WTA-CAE for reconstruction, which corresponds to multi-task learning with a self-supervised task, is denoted by S-MTL.
For both strategies it is hypothesized that, since the features learned by the WTA-CAE are related to the target labels, the additional information gained from the unlabeled data will produce a more robust model with improved generalization capability. 

In this work, F-Net training from scratch, i.e. with randomly initialized weights, was adopted as a baseline and compared to S-TL and S-MTL training. Training was conducted with a gradually increasing number of labeled samples, in order to observe the effect of limited labeled data on baseline, S-TL, and S-MTL segmentation performance. S-TL was systematically tested with a scheme of freezing, fine-tuning, and retraining of the different layers. S-MTL employed tied weights and a combined loss function for F-Net and WTA-CAE training. The approaches were investigated by using 3D liver segmentation in contrast-enhanced CT as an example.

This paper is structured as follows. Section \ref{sec:data} describes the data that was used. Section \ref{sec:core_networks} provides a general description of the two main networks, F-Net and WTA-CAE, and describes how the optimal network parameters were determined. Section \ref{sec:proposed_strategies} explains the proposed S-TL and S-MTL training strategies and the design choices. The results of the different training strategies are compared in \ref{sec:results}. Finally, section \ref{sec:discussion} discusses the results in the context of related work. 

\section{Data} \label{sec:data}
Two independent CT datasets with liver annotations and several other unannotated datasets were used for this study. Of the annotated datasets, one was used for model training and hyperparameter tuning, while the other was reserved for evaluation of the final models. 

\subsection{Annotated data}
\textbf{Training set:} $N=131$ liver training images from the Medical Segmentation Decathlon (MSD) were used. The data originated from numerous clinical sites in several countries \citep{simpson2019large}. Pathological cases included various primary cancers and metastases such as hepatocellular carcinoma and colorectal, breast, and lung cancer metastatic to the liver. Scans comprised pre- and post-treatment images and some contained metal artifacts, making for a diverse dataset representative of real-world clinical scenarios. Image resolutions ranged from $0.5-1.0$ mm in-plane, and slice thickness of $0.45 - 6.0$ mm. 

\textbf{Test set:} This set consisted of 30 images from the ``Beyond the Cranial Vault'' segmentation challenge \citep{synapseData}. Images were derived from various clinical scanners across the Vanderbilt University Medical Center (Nashville, TN, USA) and contained abnormal cases. Image voxel sizes ranged from  $0.6 - 0.9$ mm in-plane, and slice thickness of $0.5 - 5.0$ mm \citep{xu2016evaluation}.

\subsection{Unannotated data}
Supplementary data was used to further investigate the impact of a larger unlabeled to labeled data ratio on the proposed learning strategies. A total of 1277 images was available, comprising lung, pancreas, hepatic vessel, spleen, and colon train and test data from the MSD challenge. Data from these regions were included because the liver is visible in the images. Details of each dataset have been described by \citep{simpson2019large}.

\subsection{Data partitioning}
\label{sec:data_splits}
Unless stated otherwise, segmentation experiments were conducted with 5-fold cross-validation \citep{sraschka2018} using training subset $\mathcal{D}_{Tr}$ ($N_{Tr} = 105$) and validation subset $\mathcal{D}_{V}$ ($N_{V} = 26 $). In order to have each image appear once for validation, the last fold was split differently ($N_{Tr} = 104$, $N_{V} = 27$). The validation sets were disjoint across cross-validation iterations. The composition of $\mathcal{D}_{V}$ for the $k^{\rm th}$ iteration,  $k \in \{1, 2, 3, 4, 5\}$, did not vary across experiments, i.e. $\mathcal{D}_{V}$ always comprised the same images for the $k^{\rm th}$ iteration regardless of the experiment. This was done to facilitate a fair comparison between validation results of different experiment types.

$\mathcal{D}_{Tr}$ was divided into a labeled part $L_{Tr}$ and an unlabeled part $U_{Tr}$ to imitate unlabeled samples. To assess the impact of the proposed training strategies for a varying number of annotated samples, $L_{Tr}$ was generated with subsets of $n$ labeled images with $ n \in \{ 2, 4, 8, 16, 32, 64, 105 \}$ and the remaining $(N_{Tr} - n)$ images allocated to $U_{Tr}$. Within each cross-validation split, the larger subset contained the samples of the smaller subset, i.e. $L_{Tr}^{n=105} \supset L_{Tr}^{n=64} \supset L_{Tr}^{n=32}$ etc. Experiments investigating the benefit of a large number of unlabeled training images included the additional unannotated CT data in $U_{Tr}$.

\section{Core networks} \label{sec:core_networks}
This section describes the two main networks used in this study, F-Net and WTA-CAE. The network architectures are described and the training, inference, and evaluation schemes are explained. Furthermore, the experiments used to determine the optimal network parameters are presented, along with their outcomes.
All methods were implemented using the PyTorch \citep{paszke2017automatic} deep learning framework. WTA-CAE and F-Net training and inference were performed on a dedicated GPU.

\begin{figure*}[h!]
\centering
\includegraphics[width=1.00\textwidth]{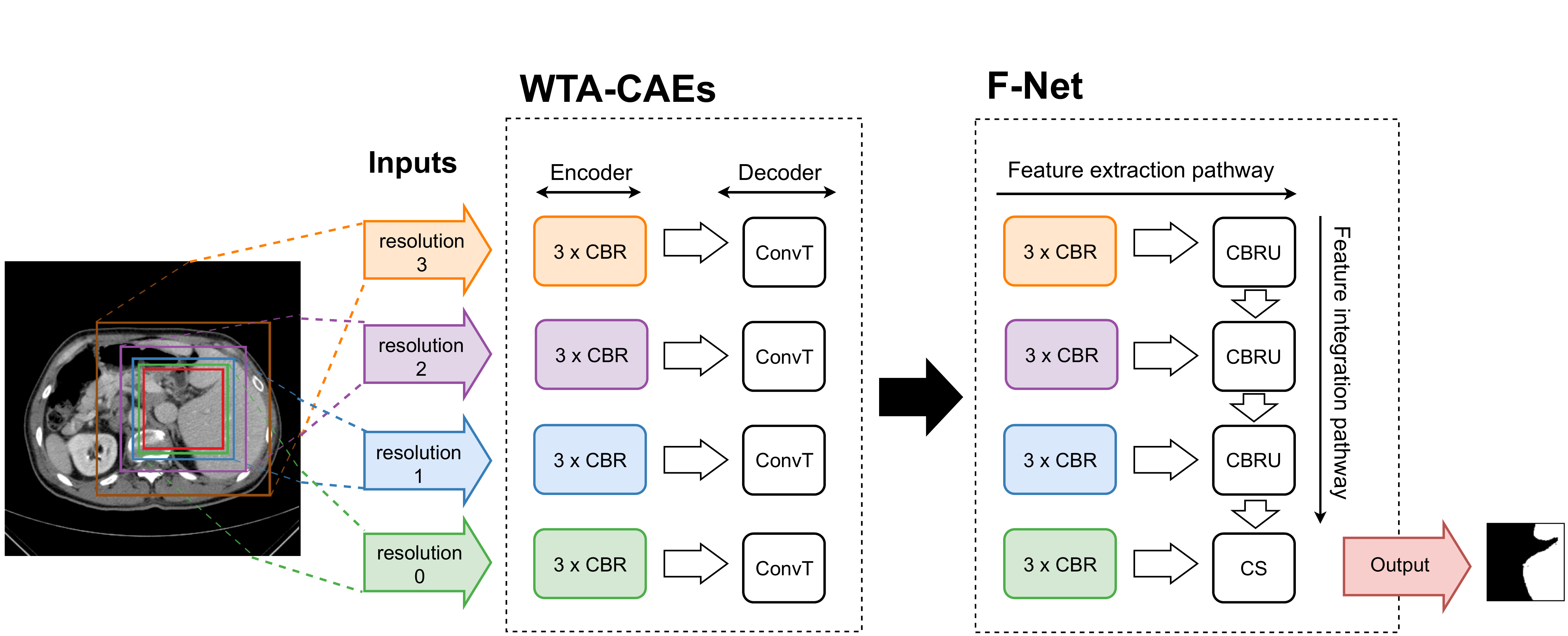}
\caption{Main network architectures used for all experiments. In order to segment the target patch (red), larger, downsampled patches at the same position are fed as input to the F-Net. Here, the input consists of four different image resolutions. 
The F-Net extracts features from the input in a feature extraction pathway and combines them again in a feature integration pathway to arrive at the final prediction.
For the proposed transfer and multi-task learning experiments an autoencoder was either pretrained for or trained together with (some of) the resolution levels of the F-Net feature extraction pathway.
The blocks denote the following operations: convolution + batch normalization + ReLU (CBR), CBR + upsampling (CBRU), convolution + softmax (CS), and transposed convolution (ConvT).}
\label{fig:architectures}
\end{figure*}

\subsection{F-Net architecture}
Fig. \ref{fig:architectures} includes a schematic representation of the F-net architecture. The network considers image patches at multiple resolution scales in order to arrive at the final prediction, combining local information gained from high resolutions with context from lower resolutions.
Unlike U-Net, which receives a single scale input image and creates the coarser resolution scales by downsampling within the network, F-Net directly receives the input as a multiscale pyramid of images. Here, an architecture with four resolution levels was used. Accordingly, each input sample to the network consisted of four images at the same position but downscaled for the lower resolution levels.

The input to each resolution level is processed in a feature extraction pathway. Thus, the number of feature extraction pathways is equal to the number of resolution levels in the network. Each feature extraction pathway comprises three successive blocks of valid convolution, batch-normalization \citep{ioffe2015batch}, and ReLU \citep{glorot2011deep} (CBR blocks). The outputs of the feature extraction levels are combined in a feature integration pathway through an additional CBR block followed by upsampling of the lower resolution outputs. Finally, a channel-wise softmax layer is applied to produce a segmentation probability map. 

\subsubsection{Training procedure} \label{sec:fnet-training}
The network was trained with the back-propagation algorithm to minimize the loss function. In particular, Focal loss and Dice - Cross Entropy loss were considered. Each training epoch consisted of a training and a validation step spanning, respectively, 15 and 8 minibatch iterations through the corresponding dataset.
Minibatches consisted of randomly selected input samples for a target segmentation patch size. 
To tackle the class imbalance between background and foreground, sampling was performed such that at least 30\% of the samples in a minibatch contained foreground region.

Convolution weights were initialized with values from a Xavier normal distribution (gain = 0.01) \citep{glorot2010understanding}. Adam optimizer \citep{kingma2014adam} was chosen to perform the weight updates with an initial learning rate (LR) of $3\times10^{-4}$ and $l_{2}$ weight decay of $3\times10^{-5}$. 
Automatic LR scheduling was adopted to decrease the LR when training reached a plateau.
As proposed in \citep{isensee2019breaking}, training was monitored using an exponential moving average of the training $l_{MA, train}$ and validation loss $l_{MA, val}$. If $l_{MA, train}$ did not decrease by at least $5 \times 10^{-3}$ within the last 50 epochs, the LR was reduced by $20\%$ until a minimum value of $10^{-7}$ was reached. The network was trained for a maximum of 1000 epochs. Training was terminated earlier if $l_{MA, val}$ did not decrease by at least $5 \times 10^{-3}$ within the last 60 epochs, but not before reaching the minimum LR.

\subsubsection{Inference}
Inference was performed per image and, in accordance with training, was also patch-based with identical patch size. Input samples were generated with 50\% overlap along each of the three axes. The output patches were aggregated using a weighted average with weights defined by a Gaussian kernel ($\sigma = \frac{1}{2} \times \text{patch size}$) centered at the patch centers \citep{isensee2018nnu}. Next, the predictions were binarized by assigning probabilities above $0.5$ label $1$ for foreground and all others label $0$ for background. The final prediction was resampled to the original image resolution to obtain a binary mask of the same size as the ground truth.

\subsubsection{Evaluation}
Segmentation performance was quantified by computing the Dice similarity coefficient (DSC), a similarity measure based on volumetric overlap.

\subsection{F-Net design choices}
Experiments were conducted on a single cross-validation fold. The following aspects of training were varied in order to determine the optimal F-Net setup. 

\subsubsection{Preprocessing}
Experiments were performed for $n=105$ and evaluated on the validation and test set to ensure that we have a well performing baseline method. 
Two preprocessing methods were investigated. Both strategies first cropped all images and corresponding labels to the region of non-zero pixel values. The images were all resampled to the median voxel spacing occurring in $L_{Tr}$. 
In method $1$, isotropic resampling of image spacings was performed to the largest median spacing across the three axes.
For each image, the intensity window was clipped at $[-155, 295]$ to optimize soft-tissue contrast and subsequently normalized to $[-3, 3]$. This preprocessing method was employed by the original F-Net paper \citep{brosch2018foveal}. 
Method 2, proposed by \cite{isensee2019breaking}, was tested with both anisotropic, as in the original paper, and isotropic resampling to the largest median spacing.
All intensity values occurring within the ground truth segmentation masks of $L_{Tr}$ were gathered and the intensity window of each image was clipped at the 0.5 and 99.5 percentiles of the collected foreground intensity values. Each image was then z-score normalized, i.e. normalized to have zero mean and unit variance.

Method 2 resulted in higher average DSC. In particular, the variant with isotropic voxel spacings performed slightly better on the test set. 
To establish a good performing baseline, Method 2 with isotropic resampling of the voxel spacings was adopted as preprocessing technique for all further experiments.

\subsubsection{Data augmentation}
In the context of neural networks, data augmentation is often used as a tool to boost model performance in the presence of limited training data. Experiments have, therefore, been performed to assess the benefit of data augmentation in the context of the chosen patch sampling strategy, whereby patch centers were randomly selected in the image. In these experiments, random on-the-fly translation (max. 10 pixels), rotation (max. 15$^{\circ}$), and scaling with log of the scaling factor between $(-0.2, 0.2)$ was applied to the images during training. Augmentation parameters were randomly sampled from a uniform distribution. The experiments were performed for varying number $n$ of labeled training images ($n \in \{2, 4, \ldots,105\}$). The resulting models were evaluated on both the validation and test set resulting in a minimal improvement in DSC on $\mathcal{D}_{V}$ and $\mathcal{D}_{Ts}$ in 2 of the 7 tested cases at the cost of $30\%$ ($+5$ hours) increase in training time. Further experiments were conducted without augmentation.

\subsubsection{Parameter selection}
Experiments were performed with the training and validation sets of $n = 105$ and were evaluated on the validation set. 

\textbf{Kernel size:} convolution kernels of size ${3}^{3}$ and $5^{3}$ were used. In accordance with the original paper \citep{brosch2018foveal}, the larger kernel produced better results for the liver segmentation task.

\textbf{Patch size:} target segmentation patch sizes of $64^{3}$ and $72^{3}$ were tested. Larger patches resulted in higher DSC.

\textbf{Minibatch size:} minibatches containing 6 and 8 samples were tested. DSC increased with more samples; 8 samples was the maximum to fit in GPU memory for target patch size $72^{3}$.

\textbf{Loss function:} F-Net was originally proposed with the Focal loss \citep{lin2017focal}. However, current experiments found a combined Dice - Cross Entropy (Dice-CE) loss to be more favorable: $\mathcal{L}_{F-Net} = \mathcal{L}_{Dice} + \mathcal{L}_{CE}$. The Dice loss formulation from \cite{isensee2018nnu} was used.
The DSC improved when directly incorporating the metric into the loss function. 

\textbf{Number of feature maps:} The original paper increased the number of feature maps for coarser resolution levels, with $[16, 24, 32, 48]$ for resolutions $0-3$. Here, the number of feature maps per level was kept constant, $[32, 32, 32, 32]$, to allow a uniform sparsity level for the autoencoders in the S-TL and S-MTL experiments. Using a uniform number of feature maps slightly lowered the average DSC, but also slightly lessened the standard deviation. The drop in performance was regarded as negligible and all further experiments used a uniform number of feature maps for the sake of the other training strategies.\\

The DSC results of the experiments are summarized in Table \ref{tab:fnet_param_tuning}. Bold results indicate the parameter choices for all further experiments.

\begin{table}[h!] 
\small
\centering
\caption{Effect of different F-Net parameter settings on the DSC (mean $\pm$ standard deviation) over the validation set ($n = 105$).}
\label{tab:fnet_param_tuning}
\begin	{tabular}{@{}lll@{}}
\toprule
Experiment                                & DSC                & DSC                \\ 
\midrule
Kernel size 3 vs. 5                       & 0.820 $\pm$ 0.082    & \textbf{0.864 $\pm$ 0.068}    \\
Patch size 64 vs. 72                      & 0.830 $\pm$ 0.076    & \textbf{0.864 $\pm$ 0.068}    \\
Minibatch size 6 vs. 8                    & 0.812 $\pm$ 0.078    & \textbf{0.864 $\pm$ 0.068}    \\
Focal loss vs. Dice-CE loss               & 0.849 $\pm$ 0.069    & \textbf{0.864 $\pm$ 0.068}    \\
\begin{tabular}[x]{@{}l@{}}Uniform vs. increasing \\ num. feat. maps\end{tabular}        & \textbf{0.864 $\pm$ 0.068}    & 0.866 $\pm$ 0.072    \\ 
\bottomrule
\end{tabular}%
\end{table}

\subsection{WTA-CAE architecture}
This autoencoder extracts features from image input with convolution operations and tries to reconstruct the input again using only the most important features. The encoder, i.e., the part of the network that extracts features, culminates in a network bottleneck. Spatial sparsity is achieved in the network by retaining only the highest activations per feature map in the bottleneck, creating a sparsity level that is proportional to the number of feature maps. Each minibatch activates different filters in the feature maps, which circumvents the dead filter problem. 
Sparsity is an essential form of regularization, since without it the network would learn useless delta functions which merely copy the input instead of extracting useful features. The decoder attempts to reconstruct the input from the sparse bottleneck through upsampling. The reconstruction is a composition of learned decoder kernels.

In this study, the encoder was identical to the F-Net feature extraction pathway and consisted of three subsequent CBR blocks. In the bottleneck resulting from the last CBR block, all except the top five highest activations in each feature map were discarded. The initially proposed 2D WTA-CAE training method \citep{makhzani2015winner} kept only the single highest activation per feature map, however, preliminary results on 3D data showed this degree of sparsity to be too extreme for learning. Using the top five activations, the network learned more effectively. As described by \cite{makhzani2015winner}, the decoder comprised a single transposed convolution layer to learn the upsampling. A simplified representation of the network architecture is shown in Fig. \ref{fig:architectures} and the correspondence between the encoder and the F-Net feature extraction pathway is illustrated. 

\subsubsection{Training procedure} \label{sec:wta-cae-training}
The training procedure was similar to that of the F-Net, with the exception of a validation step. Training was monitored through the stability of the loss function curve and visual inspection of the learned encoder and decoder convolution filters. Again, a Xavier initialization (gain = 0.01) was used for the convolution weights.

The network was trained on minibatches of image patches. Since the liver was observed to mainly be positioned in the center of the training images, patches were sampled near the center with higher frequency. This was accomplished by selecting a patch center location from the image according to a Gaussian probability sphere with its mean at the image center and $\sigma = \frac{1}{4} \times \text{image size}$. A LR scheduler, identical to that of the F-Net, monitored on $l_{MA, train}$ was used to decrease LR when training became stagnant. Training was allowed for a maximum of 500 epochs, where an epoch was defined as the iteration through 15 minibatches. 

\subsubsection{Evaluation}
The final WTA-CAE architecture and training setup were evaluated early on by applying the trained model to reconstruct a full train and test set image. The reconstruction error was confirmed to be lowest for the train image, but still reasonable for the test image, i.e., the outlines of the reconstructed structures were consistent with the input image.
Thereafter, the evaluation was performed through visual appraisal of the learned encoder and decoder weights, and the reconstructions. The best model was selected based on the largest reduction in training reconstruction error.

\subsection{WTA-CAE design choices}
The final preprocessing method was identical for the F-Net and autoencoder to ensure compatibility between the networks. Since the patch sampling was fundamentally similar to that of F-Net, data augmentation was also not used. The convolution kernel size and target segmentation patch size were chosen to be identical to the F-Net.
The following experiments were conducted on one cross-validation fold of $n=105$ to optimize the network design.

\begin{figure}[h]
\centering
\subfloat[CBR1\label{subfig:enc1}]{\includegraphics[scale=0.27]{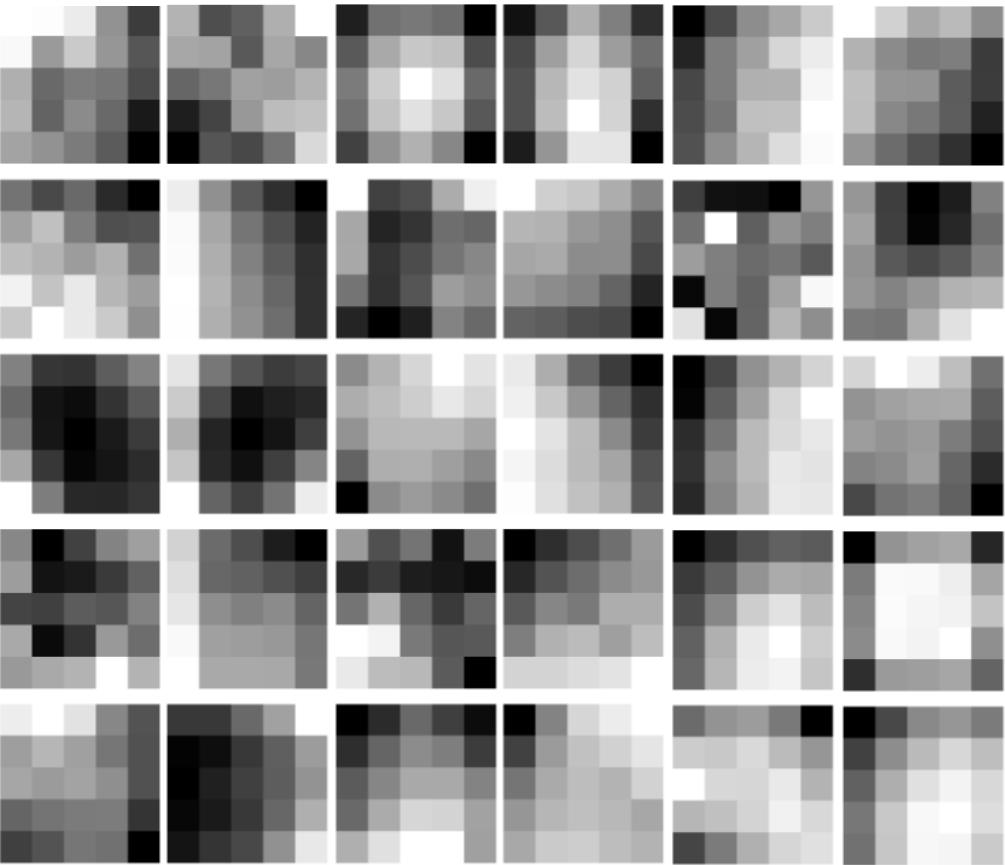}} \hspace{0.1em}
\subfloat[CBR2\label{subfig:enc2}]{\includegraphics[scale=0.27]{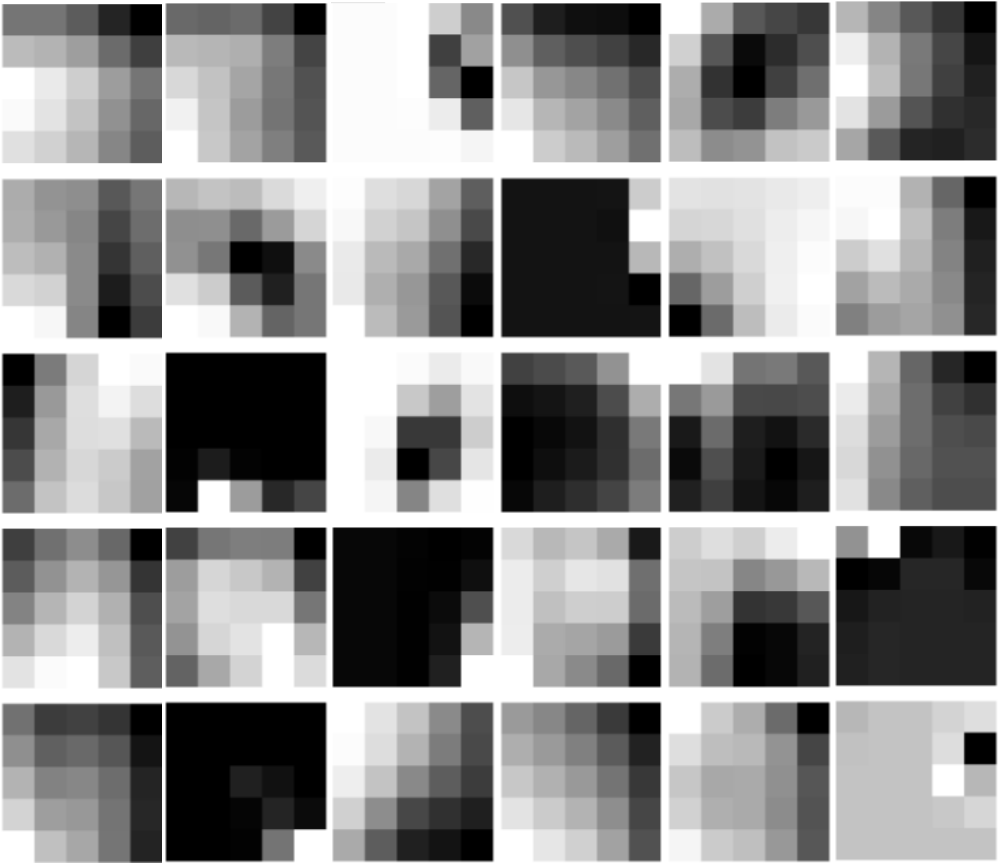}} \hspace{0.1em}
\subfloat[CBR3\label{subfig:enc3}]{\includegraphics[scale=0.27]{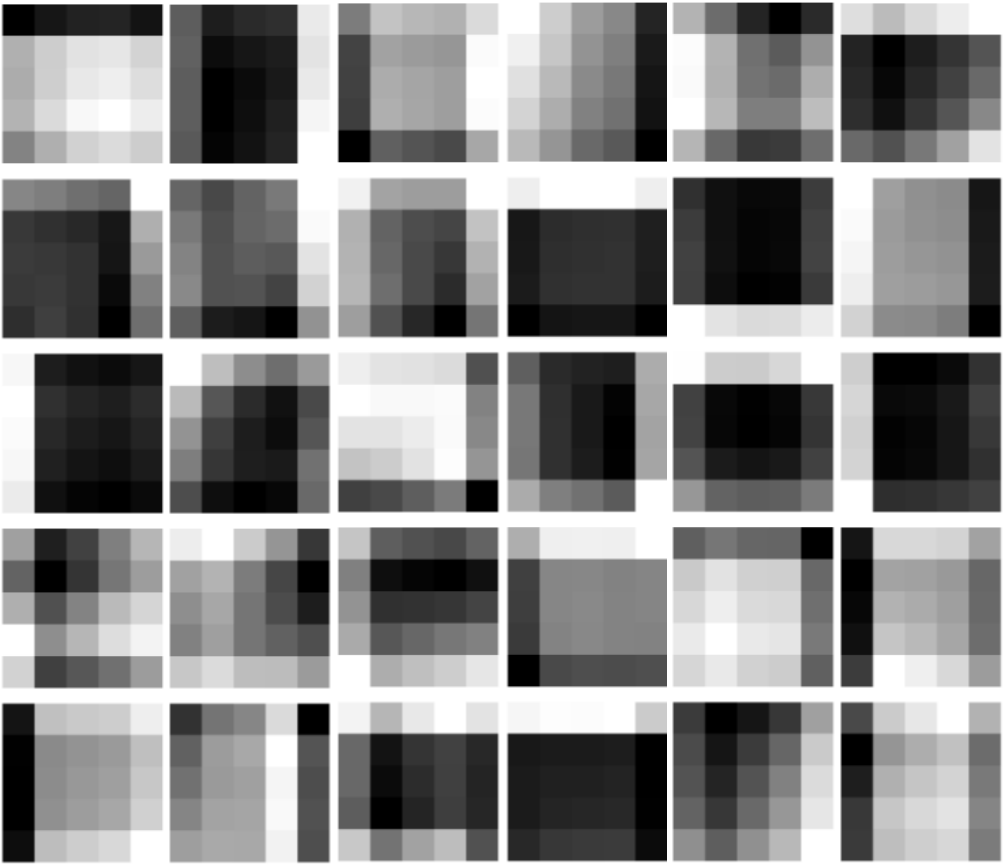}} \hspace{0.1em}
\subfloat[Dec\label{subfig:dec}]{\includegraphics[scale=0.41]{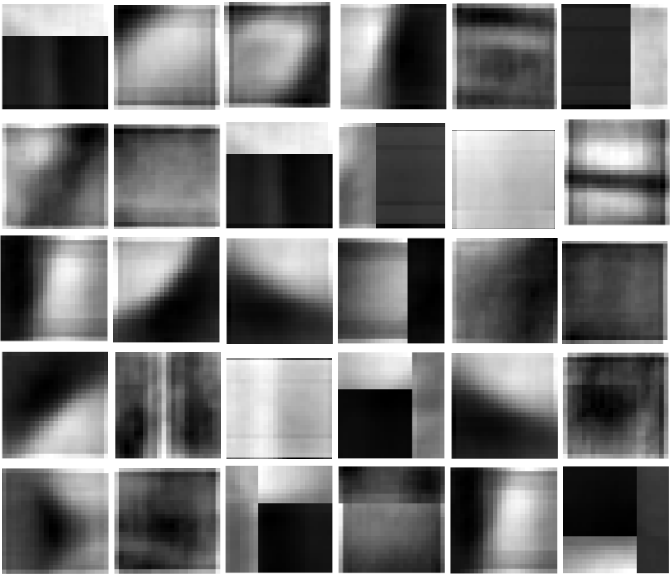}}%
\caption{Learned encoder and decoder kernels of the 3D WTA-CAE. CBR1 refers to the first convolutional block of the encoder, CBR2 to the second, etc. Dec denotes the decoder kernels.}
\label{fig:wta_kernels}
\end{figure}

\textbf{Decoder architecture:} First experiments were conducted with a decoder composed of CBR + upsampling (CBRU) block in order to closely match the F-Net feature integration pathway. This decoder did not manage to learn any sensible features, whereas a decoder comprising only a transposed convolution layer succeeded. Some representative learned kernels of the latter are visualized in Fig. \ref{fig:wta_kernels}.

\textbf{Loss function:} Preliminary experiments used an $L_{2}$ loss function for optimization, as in the original paper. The $L_{2}$ loss proved difficult to optimize since its sensitivity to outliers caused the loss function to be very noisy. Instead a Huber loss \citep{huber1992robust}, also known as smooth $L_{1}$ loss, with $\delta=1$ was used. The Huber loss is a piecewise function which equals the $L_{2}$ loss for errors smaller than 1 and $L_{1}$ loss for larger values.
The change of loss function resulted in a considerably more stable training. 

\textbf{Minibatch size:} More patches per minibatch further stabilized the training loss, even though the learned kernels were not visibly different. Therefore, the largest possible minibatch size that fits into GPU memory was used; here it was 12 patches.

\begin{figure}[h]
\centering
\subfloat[Inputs \label{subfig:input_patches}]{\includegraphics[scale=0.351]{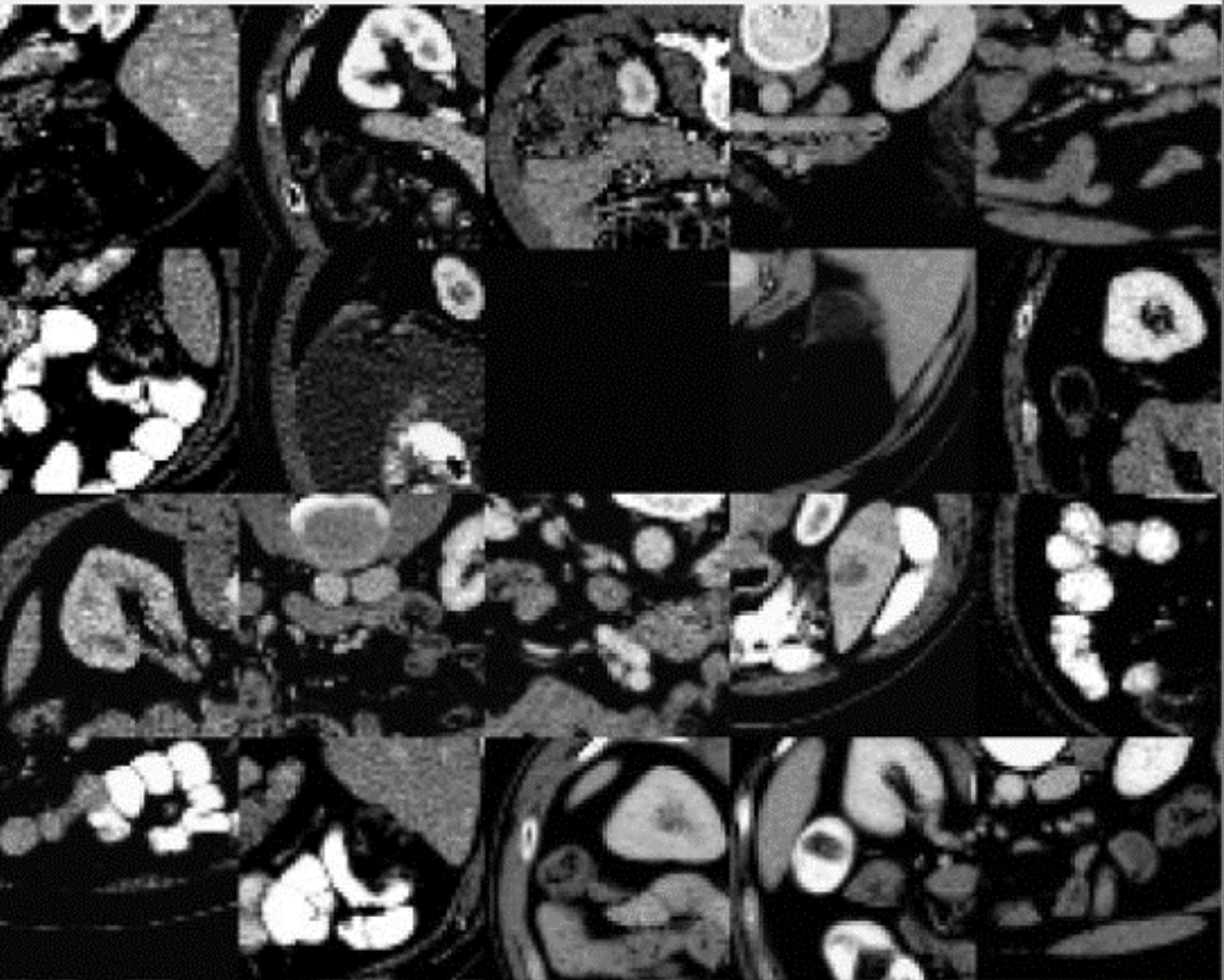}}\hspace{1.0em}
\subfloat[Reconstructions \label{subfig:recon_patches}]{\includegraphics[scale=0.9893]{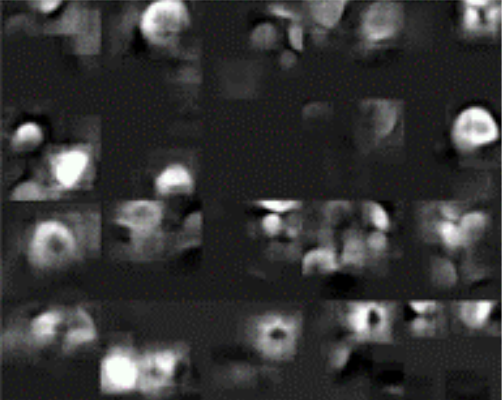}}%
\caption{Examples of learned WTA-CAE reconstructions for input patches of resolution 0.}
\label{fig:wta_recon}
\end{figure}

Exemplar reconstructions of the 3D WTA-CAE design are shown in Fig. \ref{fig:wta_recon}.

\section{Proposed learning strategies} \label{sec:proposed_strategies}
The proposed schemes of combining F-Net and WTA-CAE through transfer learning S-TL and multi-tasking learning S-MTL are described here. The experiments that motivated specific design choices are presented as well.

In all experiments, the WTA-CAE was trained on the union of images from ${L}_{Tr}$ and ${U}_{Tr}$ and F-Net was trained on ${L}_{Tr}$. F-Net and WTA-CAE training were performed according to sections \ref{sec:fnet-training} and \ref{sec:wta-cae-training} with the exception of the patch sampling method for experiments with the supplementary data. 
Since the liver's position in the unannotated images was no longer consistent, patches were uniformly sampled throughout the volume instead of more frequently near the center.
Each experiment was evaluated on the validation and test set by computing the DSC. 
As before, all methods were implemented using the PyTorch \citep{paszke2017automatic} deep learning framework and performed on a NVIDIA GTX 1080 Ti graphics card.

\begin{table*}[h!]
\centering
\caption{DSC (mean $\pm$ standard deviation) on the validation set for the various transfer learning schemes: no fine-tuning and then incremental fine-tuning of the convolution layers in the CBR blocks of the feature extraction pathway.}
\label{tab:tl_freeze_ft_scheme_val}
\resizebox{\textwidth}{!}{%
\begin{tabular}{@{}cccccc@{}}
\toprule
$n$                   & Freeze: CBR1-CBR3 & FT: CBR3 & FT: CBR2-CBR3 & FT: CBR1-CBR3 & \begin{tabular}[c]{@{}l@{}}FT: CBR2-CBR3 \\  + retrain res.3 \end{tabular}\\ \midrule
8                     & 0.901 $\pm$ 0.043 & 0.920 $\pm$ 0.033  & \textbf{0.929 $\pm$ 0.030} & 0.926 $\pm$ 0.036 & 0.921 $\pm$ 0.039          \\
16                    & 0.904 $\pm$ 0.049 & 0.921 $\pm$ 0.040  & \textbf{0.928 $\pm$ 0.040} & 0.908 $\pm$ 0.038 & 0.921 $\pm$ 0.038          \\
32                    & 0.917 $\pm$ 0.043 & 0.938 $\pm$ 0.033  & \textbf{0.943 $\pm$ 0.025} & 0.943 $\pm$ 0.027 & 0.941 $\pm$ 0.030          \\
64                    & 0.905 $\pm$ 0.037 & 0.931 $\pm$ 0.031  & 0.938 $\pm$ 0.028          & 0.936 $\pm$ 0.027 & \textbf{0.939 $\pm$ 0.025} \\
105                   & 0.913 $\pm$ 0.035 & 0.930 $\pm$ 0.035  & \textbf{0.939 $\pm$ 0.031} & 0.929 $\pm$ 0.033 & 0.939 $\pm$ 0.032          \\
\bottomrule
\end{tabular}%
}
\end{table*}

\begin{table*}[h!]
\centering
\caption{DSC (mean $\pm$ standard deviation) on the test set for the various transfer learning schemes: no fine-tuning and then incremental fine-tuning of the convolution layers in the CBR blocks of the feature extraction pathway.}
\label{tab:tl_freeze_ft_scheme_ts}
\resizebox{\textwidth}{!}{%
\begin{tabular}{@{}cccccc@{}}
\toprule
$n$                   & Freeze: CBR1-CBR3 & FT: CBR3 & FT: CBR2-CBR3 & FT: CBR1-CBR3 &  \begin{tabular}[c]{@{}l@{}}FT: CBR2-CBR3 \\   + retrain res.3 \end{tabular} \\ \midrule
8                     & 0.778 $\pm$ 0.211 & 0.807 $\pm$ 0.225 & \textbf{0.841 $\pm$ 0.133} & 0.823 $\pm$ 0.211 & 0.833 $\pm$ 0.200 \\
16                    & 0.843 $\pm$ 0.095 & 0.867 $\pm$ 0.095 & 0.873 $\pm$ 0.109 & 0.787 $\pm$ 0.237 & \textbf{0.876 $\pm$ 0.094} \\
32                    & 0.860 $\pm$ 0.077 & 0.886 $\pm$ 0.092 & \textbf{0.904 $\pm$ 0.072} & 0.898 $\pm$ 0.068 & 0.899 $\pm$ 0.065 \\
64                    & 0.855 $\pm$ 0.076 & 0.894 $\pm$ 0.064 & 0.906 $\pm$ 0.061 & 0.899 $\pm$ 0.054 & \textbf{0.911 $\pm$ 0.054} \\
105                   & 0.855 $\pm$ 0.080 & 0.891 $\pm$ 0.057 & \textbf{0.912 $\pm$ 0.049} & 0.890 $\pm$ 0.058 & 0.909 $\pm$ 0.052 \\
\bottomrule  \end{tabular}%
}
\end{table*}
\begin{table}[h!]
\small
\centering
\caption{Effect on DSC (mean $\pm$ standard deviation) when including validation for transfer learning ($n = 8$).}
\label{tab:tl_diff_datasets}
\begin{tabular}{@{}lll@{}}
\toprule
Experiment   & DSC $\mathcal{D}_{V}$ & DSC $\mathcal{D}_{Ts}$ \\ \midrule
S-TL with $\mathcal{U}_{Tr}$                                 & \textbf{0.931 $\pm$ 0.035}    &  0.839 $\pm$ 0.179   \\
S-TL with $\mathcal{U}_{Tr}$ and $\mathcal{D}_{V}$           & 0.929 $\pm$ 0.030    &  \textbf{0.841 $\pm$ 0.133}   \\
\bottomrule
\end{tabular}%
\end{table}%

\subsection{Transfer learning strategy}
Transfer learning was performed by initializing the convolutional layers of the F-Net feature extraction pathway with trained convolution weights from the WTA-CAE encoder. This required a separate WTA-CAE to be trained for each resolution level of the feature extraction pathway. Since the architectures of the feature extraction pathway and the encoders were identical, the weight transfer appeared straightforward. However, it was discovered from early experiments that the learned encoder weights differed in order of magnitude from the other F-Net weights, which led to very poor results. Therefore, a scaling factor $\phi$ was introduced to obtain similar orders of magnitude, where the weight transfer is defined as
$\mathbf{W}_{F-Net} = \phi \cdot \mathbf{W}_{WTA-CAE}$.
This strategy is depicted in Fig. \ref{fig:architectures} as a transfer of corresponding pretrained blocks from the WTA-CAEs to the F-Net.

\subsection{Design choices for transfer learning}\label{sec:tl_design_choices}
The following aspects of transfer learning were tested for one cross-validation fold. Training a single model took 26 hours on average.

\textbf{Freezing, fine-tuning, and retraining:} Generally, the earlier layers of a CNN learn common, low-level image features, whereas the later layers learn detailed, task specific features. For most applications of transfer learning, fine-tuning only the later pretrained layers is often considered sufficient for computer vision tasks. However, as shown by \cite{tajbakhsh2016convolutional} layer-wise fine-tuning experiments are necessary in order to determine the optimal transfer procedure for a specific application. Inspired by their approach, several transfer learning procedures were investigated here as well. 

For each resolution level, the convolutional layers of the transferred CBR blocks were incrementally fine-tuned from deep to shallow. Of the three CBR blocks in the F-Net feature extraction pathway, each convolutional layer was fine-tuned (FT) starting with the last block (CBR3) and including one more until all blocks (CBR1-CBR3) were being fine-tuned. Fine-tuning was performed with a low LR of $8\times10^{-5}$ to only allow small weight updates. Furthermore, an experiment was performed without FT where the pretrained weights were tested off-the shelf, i.e. all transferred weights were kept frozen during training. Experiments were conducted with $n = \{8, 16, \ldots, 105\}$.

As can be seen from the results in Tables \ref{tab:tl_freeze_ft_scheme_val} and \ref{tab:tl_freeze_ft_scheme_ts}, FT of the last two CBR blocks gave the highest DSC.
For this case, it was additionally tested if retraining the coarsest resolution, i.e. resolution 3, entirely (FT CBR2-3 + retrain res. 3) would also be favorable. However, the variant without retraining performed better overall and no further retraining experiments were performed. All remaining S-TL experiments were conducted with fine-tuning of the last two CBR blocks of each resolution level.

\textbf{Inclusion of validation images for training:} The outcome of including validation images $\mathcal{D}_{V}$, in addition to ${U}_{Tr}$, for WTA-CAE training was investigated for $n = 8$. In this case, ${U}_{Tr}$ contains $105 - 8$ images and including $\mathcal{D}_{V}$ adds $26$ images, but may also bias the validation results on $\mathcal{D}_{V}$, because the validation images have been seen during training. The evaluation was performed by computing the average DSC on the validation and test set. The 
results are shown in Table \ref{tab:tl_diff_datasets}.


The highest DSC on $\mathcal{D}_{V}$ was obtained when training with ${U}_{Tr}$ only, while for the test set $\mathcal{D}_{Ts}$ including $\mathcal{D}_{V}$ images resulted in the highest DSC. Further S-TL experiments were performed with WTA-CAEs trained on ${U}_{Tr}$ + $\mathcal{D}_{V}$, because results on the test set $\mathcal{D}_{Ts}$ may improve, but are not biased by this choice, and no data would have been available for WTA-CAE training for $n = 105$ otherwise. This approach may, however, favor S-TL over the baseline for the validation results on $\mathcal{D}_{V}$ and needs to be taken into account when interpreting the results. 

\begin{table*}[h!]
\small
\centering
\caption{DSC (mean $\pm$ standard deviation) on the validation and test set for S-MTL training with WTA-CAEs for a different number of resolution levels.}
\label{tab:mtl_effect_levels}
\begin{tabular}{c@{\qquad}ccccc}
  \toprule
  & $n$ & res. $0$ & res. $0 - 1$ & res. $0 - 2$ & res. $0 - 3$ \\  
  \midrule
	\multirow{2}{*}{\raisebox{-\heavyrulewidth}{$\mathcal{D}_{V}$}} 
  & 8                     & 0.922 $\pm$ 0.036 & \textbf{0.931 $\pm$ 0.032} & 0.928 $\pm$ 0.034 & 0.926 $\pm$ 0.035 \\
  & 105                   & 0.939 $\pm$ 0.034 & 0.940 $\pm$ 0.029 & \textbf{0.941 $\pm$ 0.036} & 0.936 $\pm$ 0.032 \\ 
	\midrule
	\multirow{2}{*}{\raisebox{-\heavyrulewidth}{$\mathcal{D}_{Ts}$}} 
  & 8                     & 0.853 $\pm$ 0.107 & \textbf{0.872 $\pm$ 0.081} & 0.856 $\pm$ 0.101 & 0.851 $\pm$ 0.103 \\
  & 105                   & 0.899 $\pm$ 0.059 & \textbf{0.904 $\pm$ 0.052} & 0.903 $\pm$ 0.054 & 0.889 $\pm$ 0.062 \\
  \bottomrule  
\end{tabular}%
\end{table*}

\begin{table}[h!]
\small
\centering
\caption{Effect on DSC (mean $\pm$ standard deviation) when including validation images for S-MTL ($n = 8$).}
\label{tab:mtl_diff_datasets}
\begin{tabular}{@{}lll@{}}
\toprule
Experiment   &  DSC $\mathcal{D}_{V}$ & DSC $\mathcal{D}_{Ts}$ \\ \midrule
S-MTL with $\mathcal{U}_{Tr}$                                 & 0.917 $\pm$ 0.037                & 0.851 $\pm$ 0.099        \\
S-MTL with $\mathcal{U}_{Tr}$ and $\mathcal{D}_{V}$           & \textbf{0.931 $\pm$ 0.032}       & \textbf{0.872 $\pm$ 0.081}        \\
\bottomrule
\end{tabular}%
\end{table}%

\subsection{Multi-task learning strategy}
Mutli-task learning was performed by training the F-Net and WTA-CAEs in parallel for their individual tasks. Training procedures were in line with sections \ref{sec:fnet-training} and \ref{sec:wta-cae-training}, except for the minibatch sizes. Issues with GPU memory required F-Net and WTA-CAE minibatch sizes to be reduced to 5 and 8, respectively. 

Again, a distinct WTA-CAE was trained for every F-Net resolution level. Per resolution level, the corresponding convolutional layers of the feature extraction pathway and of the encoder were shared, i.e., tied weights were used. Weight tying significantly reduced the number of learnable parameters in the network and enabled training of the two networks in parallel. Corresponding blocks are depicted with similar colors in Fig. \ref{fig:architectures}. 
The feature integration pathway and the different decoders remained independent of each other, allowing each network to retain their own optimizer and loss. The final loss function $\mathcal{L}_{MTL}$ was the sum of the F-Net loss and the average loss of the WTA-CAEs. The latter was computed as the arithmetic mean of the four WTA-CAE losses with a scaling factor $\gamma$:
$\mathcal{L}_{MTL} = \mathcal{L}_{F-Net} + \gamma \cdot \mathcal{L}_{WTA-CAE}^{avg}$

\subsection{Design choices for multi-task learning}
The experiments were conducted on a single cross-validation fold. One model took roughly 39 hours to train.

\textbf{Loss scaling factor $\gamma$:} First $\gamma$ was calculated so that $\mathcal{L}_{F-Net}$ and $\mathcal{L}_{WTA-CAE}^{avg}$ had similar orders of magnitude. Experiments were then performed for $n=8$ and $105$, where the contribution of $\mathcal{L}_{WTA-CAE}^{avg}$ was varied by multiplying $\gamma$ by an additional factor $f: \{0.1, 0.2, \ldots, 1.0\}$. 

There was no clear link between $f$ and the obtained DSC results for any $n$ and it was unclear if the minor differences between results for different $f$ were merely due to random aspects of training. For this reason, further experiments were performed with $f=1.0$ so that $\gamma$ only equalized the orders of magnitude.

\textbf{Excluding levels for MTL:} 
To determine the utility of training a WTA-CAE in parallel for each F-Net resolution level, experiments were conducted where a WTA-CAE was trained for only a single level (resolution 0), and then added for consecutive levels until there was a WTA-CAE for each F-Net resolution level. The experiments were conducted with $n=8$ and $105$. 

The average DSC of these experiments on the validation and test set are presented in Table \ref{tab:mtl_effect_levels}. 
Almost all cases achieved the highest DSC when training only the first two resolution levels. Only $n=105$ achieved the highest average DSC on the validation set by training three resolution levels. However, training the first two resolutions for this case achieved the second highest DSC with the smallest standard deviation.
Therefore, training a WTA-CAE for only the first two F-Net resolutions was considered the best approach to S-MTL and used for all remaining experiments.

\textbf{Inclusion of validation images for training:} 
The experimental setup was identical to that of S-TL described in section \ref{sec:tl_design_choices}. The results are presented in Table \ref{tab:mtl_diff_datasets}. 
The addition of $\mathcal{D}_{V}$ images clearly improved results on both the validation and test datasets. Though a similar improvement can be observed in both cases, it may be attributed for $\mathcal{D}_{V}$ to the fact that the validation images have been seen during training. In order to be consistent with S-TL, further S-MTL experiments were performed with WTA-CAEs trained on ${U}_{Tr}$ + $\mathcal{D}_{V}$. Again, validation results on $\mathcal{D}_{V}$ may favor S-MTL over the baseline which must be taken into account when interpreting the results. 

\section{Results}\label{sec:results}
This section presents and compares the DSC results of baseline, i.e., F-Net with randomly initialized weights, S-TL, and S-MTL training.
Training was performed for all $n: \{2, 4, \ldots,105\}$ specified in section \ref{sec:data_splits}, except for the S-TL and S-MTL experiments using the supplementary unannotated data. Experiments using the additional data were performed for $n = 32$ and $105$.
Each experiment for baseline, transfer learning, and multi-task learning is the result of five cross-validation instances and took, respectively, 80 hours, 130 hours and 195 hours to compute.

\begin{figure*}[h!]
\centering
\includegraphics[width=\textwidth]{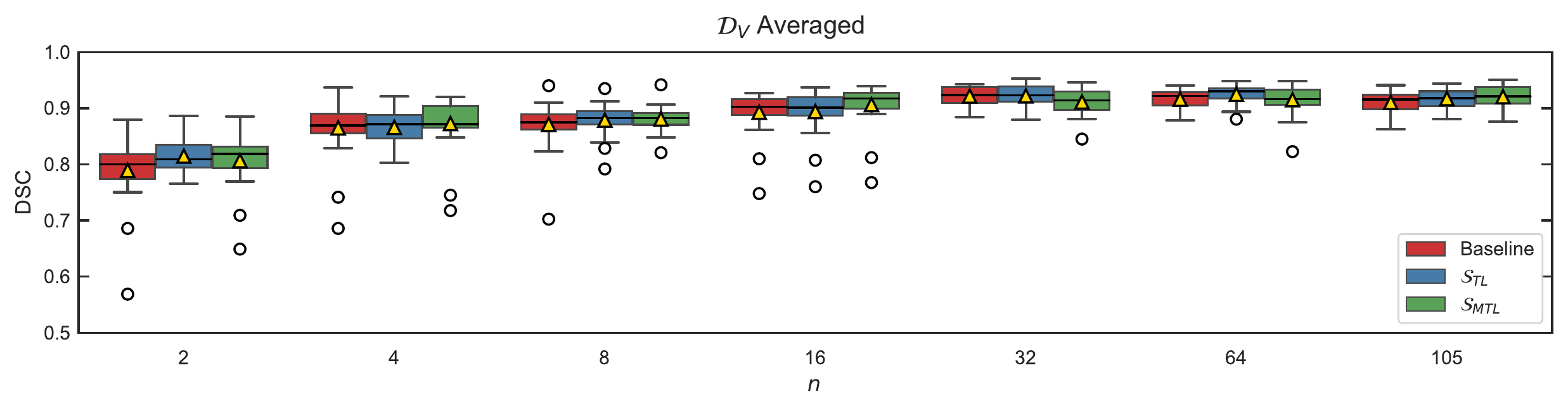}%
\caption{DSC of the five cross-validation models on the validation set. Mean values are denoted by a yellow triangle. Whiskers represent $1.5 \times$ IQR. Note that the y-axis does not start at zero.}
\label{fig:res_V_avg}
\centering
\includegraphics[width=\textwidth]{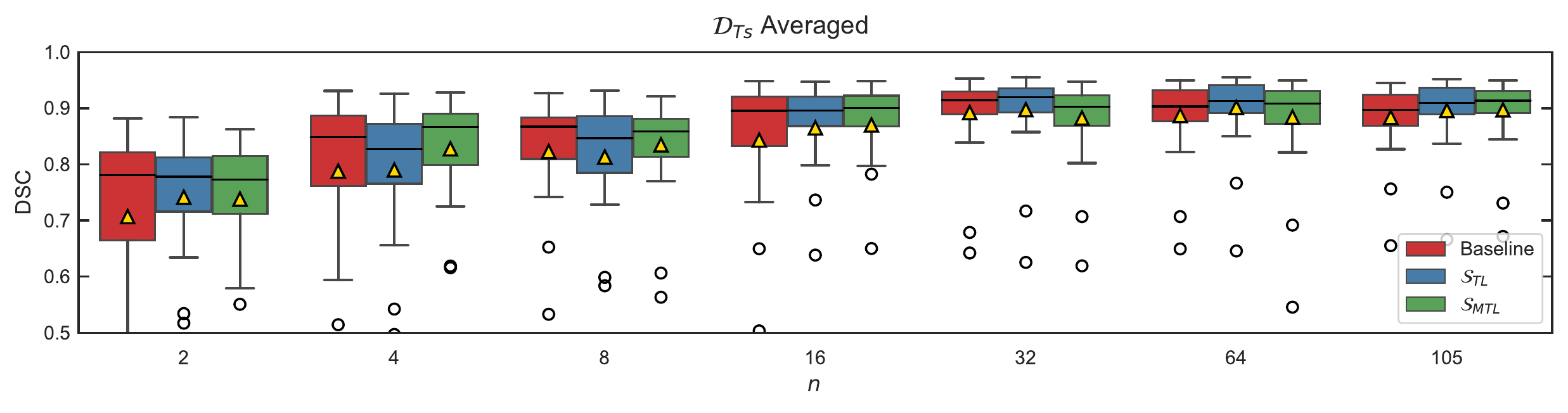}%
\caption{DSC of the five cross-validation models on the test set. Mean values are denoted by a yellow triangle. Whiskers represent $1.5 \times$ IQR. Note that the y-axis does not start at zero.}
\label{fig:res_Ts_avg}
\end{figure*}

\begin{figure}[h!]
  \centering
  \includegraphics[scale=0.65]{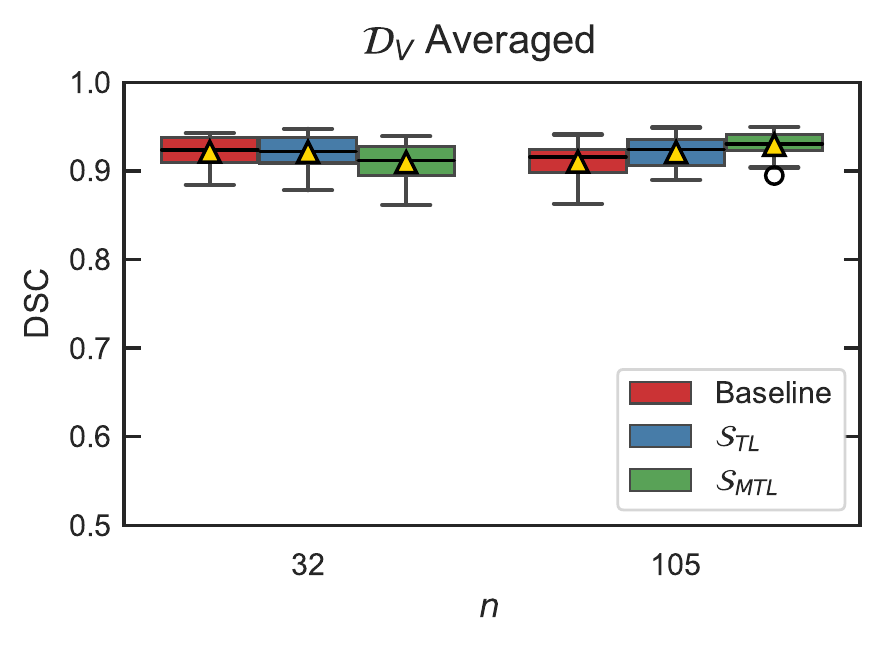}
  \caption{DSC of the five cross-validation models validation set, with the supplementary data set included for training. Mean values are denoted by a yellow triangle. Whiskers represent $1.5 \times$ IQR. Note that the y-axis does not start at zero.}
  \label{fig:res_V_other_avg}
\end{figure}

\begin{figure}[h!]
  \centering
  \includegraphics[scale=0.65]{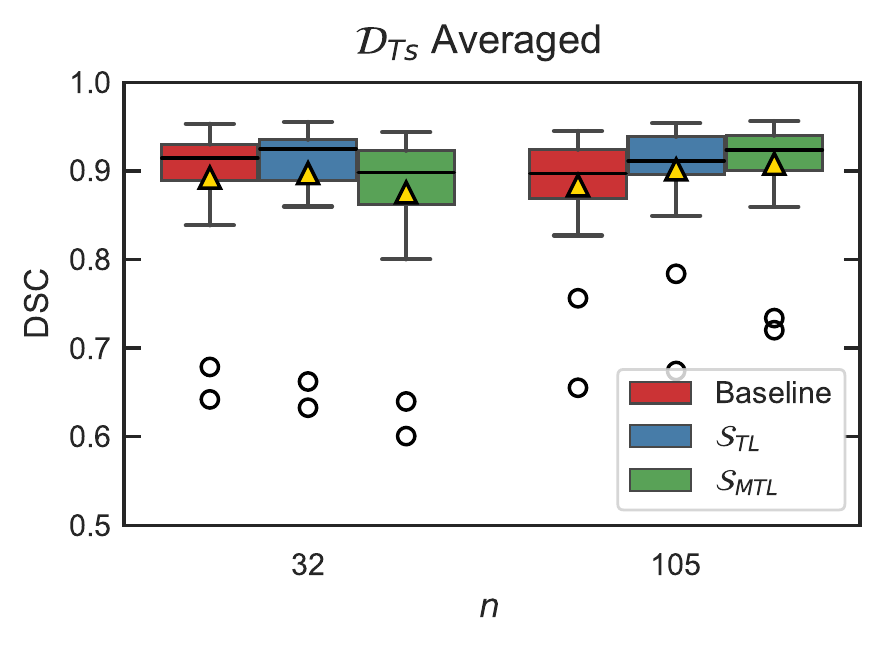}
  \caption{DSC of the five cross-validation models test set, with the supplementary data set included for training. Mean values are denoted by a yellow triangle. Whiskers represent $1.5 \times$ IQR. Note that the y-axis does not start at zero.}
  \label{fig:res_Ts_other_avg}
\end{figure}

Boxplots of the DSC of each training strategy are displayed alongside each other in Fig. \ref{fig:res_V_avg} (validation set) and \ref{fig:res_Ts_avg} (test set) for the different $n$. The results of the experiments using the supplementary data are shown in Fig. \ref{fig:res_V_other_avg} (validation set) and \ref{fig:res_Ts_other_avg} (test set). These boxplots reflect the spread in DSC within the validation and test datasets.

To allow for a comparison of the numerical values, the DSC results averaged over the cross-validation folds on the validation set and the test set, without supplementary data, are presented in Table \ref{tab:different_strategies_results}. The results using the additional data are shown in Table \ref{tab:different_strategies_results_additional_data}. 

\begin{figure}[h!]
\centering
\includegraphics[scale=0.73]{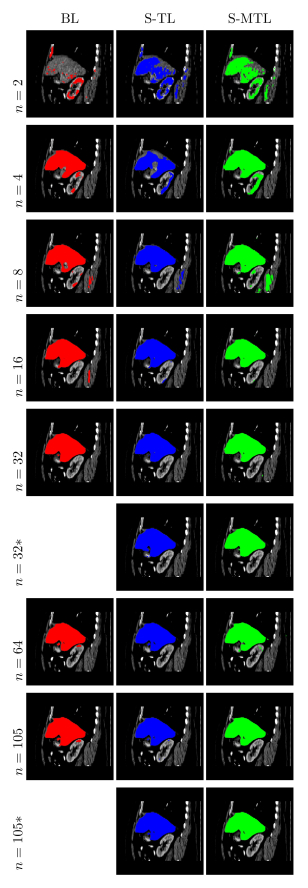}
\caption{Sagittal segmentations of an exemplar test image for baseline (BL), S-TL, and S-MTL with an increasing number of labeled training images ($n$). $*$ indicates that the additional unlabeled data were included for training.}
\label{fig:exemplar_sagittal_segmentations}
\end{figure}

Exemplar segmentations of a test image for the different training strategies are shown in Fig. \ref{fig:exemplar_sagittal_segmentations}.

\textbf{Baseline:} As expected, baseline results improved by adding more labeled training data, up to $n=32$. After $n=32$ this trend ceased. The results for $n=32$ stand out since the DSC is higher than all other cases. The reason for this is not well understood, but it can be ruled out that it is the result of an unfortunate data division for training and validation. Special care was taken to include each smaller subset of $n$ in the next larger training subset (section \ref{sec:data_splits}). Moreover, the results are the average of training and validation with five different data splits. $n=32$ is therefore regarded as an outlier. 

\textbf{Transfer learning:} 
Compared to baseline, the proposed transfer learning strategy slightly increased the DSC on the validation set in 7 out of 9 tested cases and on the test set in 8 of 9 cases. 
The largest improvement of $0.034$ was observed for $n = 2$ from a baseline DSC of $0.707$ to $0.741$ for the test set.
From $n=2$ up to $64$, the DSC increased as $n$ grew larger. Notably, the S-TL experiments without supplementary unannotated data still improved upon baseline for $n=105$, despite the number of new unlabeled images being reduced as $n$ increases, i.e. just $26$ new images for $n=105$.
As can be seen in Table \ref{tab:different_strategies_results_additional_data}, using a much larger set of unlabeled training data gave further improvement in DSC. Again the case of $n=32$ formed an exception as the result on the validation set was slightly poorer. 
For $n=105$, S-TL using additional unlabeled data improved the baseline DSC from $0.910$ to $0.921$ and $0.884$ to $0.902$ for validation and test set, respectively.
Fig. \ref{fig:res_Ts_avg} and \ref{fig:res_Ts_other_avg} show several cases where the transfer learning strategy reduces the spread in DSC values compared to baseline.

\textbf{Multi-task learning:} 
The proposed multi-task learning strategy slightly improved the DSC on both validation and test set in 6 out of 9 cases.
The largest improvement of $0.040$ was observed for $n = 4$ from a baseline DSC of $0.788$ to $0.828$ for the test set.
Table \ref{tab:different_strategies_results} shows that the DSC increased as $n$ became larger, up to $n=105$. 
As with transfer learning, note that S-MTL still improves upon baseline for $n=105$ despite having only a small number of new images available for the WTA-CAE.
Again, Table \ref{tab:different_strategies_results_additional_data} shows that using much more unlabeled data further improved the DSC on the validation and test set. 
The improvement is especially apparent for the $n=105$ case, where S-MTL improved the baseline DSC from $0.910$ to $0.930$ and $0.884$ to $0.908$ for validation and test set, respectively.
As can be seen from Fig. \ref{fig:res_Ts_avg} and \ref{fig:res_Ts_other_avg}, multi-task learning also reduces the spread in DSC values in several cases compared to baseline.


\begin{table*}[h!]
\centering
\caption{DSC results (mean $\pm$ standard deviation) of the different learning strategies on the validation and test set. Note that as $n$ increases, the number of new unlabeled samples ($N - n$) for the WTA-CAE decreases, since no additional unlabeled data was used here and the training set size is limited to $N$.}
\label{tab:different_strategies_results}
\resizebox{\textwidth}{!}{%
\begin{tabular}{c@{\qquad}ccc@{\qquad}ccc}
  \toprule
  \multirow{2}{*}{\raisebox{-\heavyrulewidth}{$n$}} & \multicolumn{3}{c}{Validation set $\mathcal{D}_{V}$} & \multicolumn{3}{c}{Test set $\mathcal{D}_{Ts}$} \\
  \cmidrule{2-7}
  & Baseline & S-TL & S-MTL & Baseline & S-TL & S-MTL\\
  \midrule
  $2$   & 0.789 $\pm$ 0.058 & 0.815 $\pm$ 0.028 & 0.807 $\pm$ 0.045 & 0.707 $\pm$ 0.164 & 0.741 $\pm$ 0.116 & 0.738 $\pm$ 0.114 \\
  $4$   & 0.865 $\pm$ 0.051 & 0.866 $\pm$ 0.029 & 0.873 $\pm$ 0.046 & 0.788 $\pm$ 0.165 & 0.790 $\pm$ 0.130 & 0.828 $\pm$ 0.104 \\
  $8$   & 0.871 $\pm$ 0.041 & 0.879 $\pm$ 0.028 & 0.881 $\pm$ 0.022 & 0.823 $\pm$ 0.111 & 0.813 $\pm$ 0.109 & 0.835 $\pm$ 0.080 \\
  $16$  & 0.893 $\pm$ 0.038 & 0.894 $\pm$ 0.038 & 0.906 $\pm$ 0.037 & 0.843 $\pm$ 0.133 & 0.865 $\pm$ 0.106 & 0.870 $\pm$ 0.098 \\
  $32$  & 0.922 $\pm$ 0.018 & 0.922 $\pm$ 0.020 & 0.911 $\pm$ 0.023 & 0.892 $\pm$ 0.070 & 0.897 $\pm$ 0.068 & 0.883 $\pm$ 0.069 \\
  $64$  & 0.916 $\pm$ 0.017 & 0.925 $\pm$ 0.017 & 0.915 $\pm$ 0.026 & 0.887 $\pm$ 0.066 & 0.901 $\pm$ 0.062 & 0.884 $\pm$ 0.083 \\
  $105$ & 0.910 $\pm$ 0.021 & 0.917 $\pm$ 0.018 & 0.921 $\pm$ 0.020 & 0.884 $\pm$ 0.060 & 0.896 $\pm$ 0.060 & 0.897 $\pm$ 0.060 \\
  \bottomrule
\end{tabular}%
}
\vspace{1.0em}
\centering
\caption{DSC results (mean $\pm$ standard deviation) of the proposed learning strategies, using the additional set of unlabeled data, on the validation and test set. Note that for these experiments, the number of unlabeled training images is $1277 + (N-n)$.}
\label{tab:different_strategies_results_additional_data}
\resizebox{\textwidth}{!}{%
\begin{tabular}{c@{\qquad}ccc@{\qquad}ccc}
  \toprule
  \multirow{2}{*}{\raisebox{-\heavyrulewidth}{$n$}} & \multicolumn{3}{c}{Validation set $\mathcal{D}_{V}$} & \multicolumn{3}{c}{Test set $\mathcal{D}_{Ts}$} \\
  \cmidrule{2-7}
  & Baseline & S-TL & S-MTL & Baseline & S-TL & S-MTL\\
  \midrule
  $32$  & 0.922 $\pm$ 0.018 & 0.915 $\pm$ 0.032 & 0.915 $\pm$ 0.029 & 0.892 $\pm$ 0.070 &0.901 $\pm$ 0.069 & 0.890 $\pm$ 0.078 \\
  $105$ & 0.910 $\pm$ 0.021 & 0.921 $\pm$ 0.019 & 0.930 $\pm$ 0.015 & 0.884 $\pm$ 0.060 &0.902 $\pm$ 0.056 & 0.908 $\pm$ 0.055 \\
  \bottomrule
\end{tabular}%
}
\end{table*}

\section{Discussion} \label{sec:discussion}
In this work, the integration of knowledge from unlabeled data with a 3D CNN segmentation model was investigated. 3D Liver segmentation in CT images was used as an example. An autoencoder, trained according to the Winner-Take-All method, was used to extract features intended for input reconstruction from the unannotated medical volumes. This information was used in the training of an F-Net segmentation network through transfer and multi-task learning strategies. These two strategies were compared to F-Net training with randomly initialized weights, the baseline. Experiments were performed with larger and reduced amounts of labeled and unlabeled data, in order to determine the impact of these factors on the proposed learning strategies. It was hypothesized that a WTA-CAE could learn discriminative features from unannotated images to improve the, either subsequent or concurrent, target task of segmentation with an F-Net.

From the experiments with reduced labeled data, $n=32$ stood out. For smaller $n$, baseline results improved with the addition of more labeled samples as expected, but after $n=32$ both validation and test results worsened. It could not be explained why baseline training reached its peak with $32$ labeled samples. A possibility could be random factors in the training, but it cannot be ruled out that pathological cases in the dataset, the chosen preprocessing method or implicit choices in the training framework were the cause. Current results for $n=32$ notwithstanding, it is believed that, in general, addition of a significant amount of labeled training data improves results, and the results for $n=32$ are considered to be an outlier. 


Overall, both S-TL and S-MTL improved over the baseline training and increased the dice score by up to $0.034$ ($n = 2$), $0.040$ ($n = 4$) and $0.024$ ($n = 105$ with supplementary data), respectively. The tendencially larger improvements when using few labeled samples for training and the results for $n = 105$ without and with the supplementary unannotated data indicate that both training strategies are more effective with a large ratio of unlabeled to labeled training data which has been about $64 : 1$ ($n = 2$), $32 : 1$ ($n = 4$), and $12.5 : 1$ ($n = 105$ with supplementary unannotated data). The improvements tend to be larger for the test set than for the validation set, which indicates that validation set results are not biased by S-TL and S-MTL training. S-MTL performed slightly better than S-TL despite being forced to use smaller minibatches due to limited resources which was found to significantly degrade the F-Net segmentation performance (Table \ref{tab:fnet_param_tuning}). However, S-TL is the least computationally complex as it retains the same number of network parameters but with an extra pretraining step. 

It should be acknowledged that various parameter settings have initially been optimized for the F-Net and afterwards used for S-TL and S-MTL, mainly with the goal to have a well performing baseline and consistent results. The performance of neural networks depends, however, significantly on many details like preprocessing, parameter selection and network adaptation \citep{isensee2019breaking}. Results may, therefore, be biased towards the F-net and combined tuning of the F-Net and WTA-CAE for the proposed learning strategies might further improve results. 

The supplementary unannotated data have been acquired for different tasks (lung, pancreas, hepatic vessels, spleen, and colon). The covered anatomy was in many datasets different from that of the training, validation and test data and the uniform patch sampling strategy of the WTA-CAE supplied many input patches which did not contain relevant structures. One way to resolve this and further improve the use of the supplementary unannotated data would be to estimate the general position of the liver in the images for each task and sample patches with a Gaussian probability from this region. Nevertheless, the results for $n = 105$ show that use of these data during training was beneficial.

Other liver segmentation studies have reported higher DSC values than have been presented here. For example, \cite{isensee2018nnu}, winner of the MSD challenge, achieved a DSC of $95.43\%$ for the Decathlon liver task. These values were produced by, inter alia, cascading and ensembling multiple models. Such additional methods of improving results were not used for this study in order to more clearly see the influence of the proposed training strategies. To this end, it was also decided not to employ any post-processing techniques to remove spurious segmentations in the images. Otherwise the achieved DSC would have been higher. 

Adaptation of the proposed training strategies to the more common U-Net architecture is possible, but modifications might be needed to avoid memory issues. In addition, constructing a deeper autoencoder architecture in the image of U-Net instead of training a separate model per resolution level would be possible, but deviates from the Winner-Take-All training strategy. The method of constructing a multi-resolution, stacked autoencoder as a way to pretrain CNN weights has previously been proposed by \cite{masci2011stacked} and has achieved superior performance in classification tasks. This finding bodes well for an adaptation to U-Net using a single stacked autoencoder.

In comparing the current results to similar works found in literature, it was found that \cite{bai2019self} used remarkably similar training strategies. In their recent work, the authors also explored transfer and multi-task learning strategies with labeled and unlabeled data for 2D cardiac MR segmentation. A U-Net architecture was used with anatomical position prediction as auxiliary task. As was the case here, they found that the multi-task learning strategy most often improved U-Net training from scratch, especially for few labeled training samples. It should be noted that their study used a larger unlabeled to labeled data ratio of roughly $350:1$. 

As in this study, \cite{tajbakhsh2019surrogate} looked at self-supervised learning for 3D medical image segmentation. They used rotation prediction to pretrain a dense V-Net for lung lobe segmentation, a task which requires strong supervision and large amounts of labeled data. Their results clearly showed that pretraining with a surrogate task improved over training the network from scratch, using unlabeled to labeled data ratios of approximately $40:1$, holding up to $8:1$. Also, pretraining was more effective for smaller amounts of labeled training samples, which consequently increased their amount of unlabeled data. Their findings agree with the ones here, that self-supervised learning is most effective when there is a larger ratio of unlabeled to labeled data.

Slightly more distant from the current context, \cite{ross2018exploiting} also successfully applied self supervision for a 2D segmentation problem with endoscopic video data. The authors used image colorization as the auxiliary task, wherein colonoscopy images were converted to gray-scale and then recovered using a conditional Generative Adversarial Network. Also \cite{chen2019self} reported improved results when applying self-supervised learning for, inter alia, 2D tumor segmentation from brain MR images. They used their own novel auxiliary task of context restoration. As before, both papers reported that self-supervised learning provided the largest performance gain for small sets of labeled data.

\section{Conclusion}
This study investigated the potential of learned autoencoder features for improving 3D CNN segmentations when additional unlabeled data for the target region is available. A F-Net (3D CNN) was used as baseline and combined with a WTA-CAE autoencoder. Two strategies have been considered. The first strategy uses transfer learning and initializes the weights of an F-Net with weights of a WTA-CAE autoencoder. The second strategy uses multi-task learning and simultaneously trains an F-Net for segmentation and WTA-CAE for reconstruction. Results of systematic experiments with increasing amount of labeled training images show that the learning strategies improved results in $75\%$ of the experiments compared to training from scratch. The dice score increased by up to $0.040$ and $0.024$ for a ratio of unlabeled to labeled training data of about $32 : 1$ and $12.5 : 1$, respectively. The results indicate that both training strategies are more effective with a large ratio of unlabeled to labeled training data. Multi-task learning performed slightly better, but tends to demand more computational resources than the transfer learning strategy.

\bibliography{mybibfile}

\end{document}